\documentclass[aps,reprint]{revtex4-1}
\usepackage[dvips]{graphicx}
\usepackage{color}

\begin{document}


\title{Few Graphene layer/Carbon-Nanotube composite Grown at CMOS-compatible Temperature} 



\author{V. Jousseaume$^{1}$}
\email{vincent.jousseaume@cea.fr}
\author{J. Cuzzocrea$^{1}$}
\author{N. Bernier$^{1}$}
\author{V. T. Renard$^{1,2}$} 

\affiliation{$^{1}$CEA, LETI, Minatec, F38054 Grenoble, France\\
$^{2}$Institut N\'{e}\'{e}l, CNRS-UJF, BP 166, 38042 Grenoble, France}



\date{\today}

\begin{abstract}
We investigate the growth of the recently demonstrated composite material composed of
vertically aligned carbon nanotubes capped by few graphene layers. We show that the carbon nanotubes
grow epitaxially under the few graphene layers. By using a catalyst and gaseous carbon precursor
different from those used originally we establish that such unconventional growth mode is not specific
to a precise choice of catalyst-precursor couple. Furthermore, the composite can be grown using catalyst
and temperatures compatible with CMOS processing ($T < 450$ $^{\circ}$C). 
\end{abstract}


\maketitle 

The limitations of copper interconnects are driving research towards alternative materials and technologies 
for the next-generation of Integrated Circuits (ICs). Carbon nanomaterials, with their many attractive properties, are emerging 
as the frontrunners to potentially replace copper for interconnects and passive devices in ICs, including vias, through-silicon 
vias (TSVs) and horizontal wires \cite{Kreupl2002,Nihei2005,Awano2006,Wang2009}. Indeed, low-dimensional allotropes of carbon (in particular, carbon nanotubes (CNTs) and graphene) 
have extraordinary physical properties because of their unique structure. Contrary to copper, CNTs can accept a high current density 
and can have an extremely high thermal conductivity \cite{Yao2000,Kim2001}. Graphene has the potential for very high electrical conductivity \cite{Neugebauer2009}. 
Beside these advantages for interconnects technology, Field Effect Transistors have been demonstrated on CNTs and graphene.\cite{Tans1998,Lin2010} 
It could be very interesting to combine these advances in interconnects and transistor technologies in an \textit{all carbon} electronics. The succes of such
 approach could strongly depend on the ability to assemble them in working devices. Recently, Fujitsu Corp. has demonstrated the possibility of building 
carbon interconnect structures combining CNTs and few graphene layers\cite{Kondo2008}(these structures will we reffered to as ``composite'' in the following). This 
approach of binding CNTs and few-layer graphene during the growth itself maybe a powerfull method to produce CNT/graphene hybrid systems.

Here, we show that such composite is composed of CNTs epitaxially grown bellow the few graphene layers. We demontrate that the composite can 
be obtained independently of the choice of the precursor and catalyst indicating that this unusal growth mode is in fact quite general.
Furthermore, we achieved growth using catalyst and temperatures compatible with Back End of Line (BEOL) processing. This opens new possibilities concerning 
its integration in next generations of interconnects. 
\begin{figure}
\includegraphics[width=0.7\columnwidth]{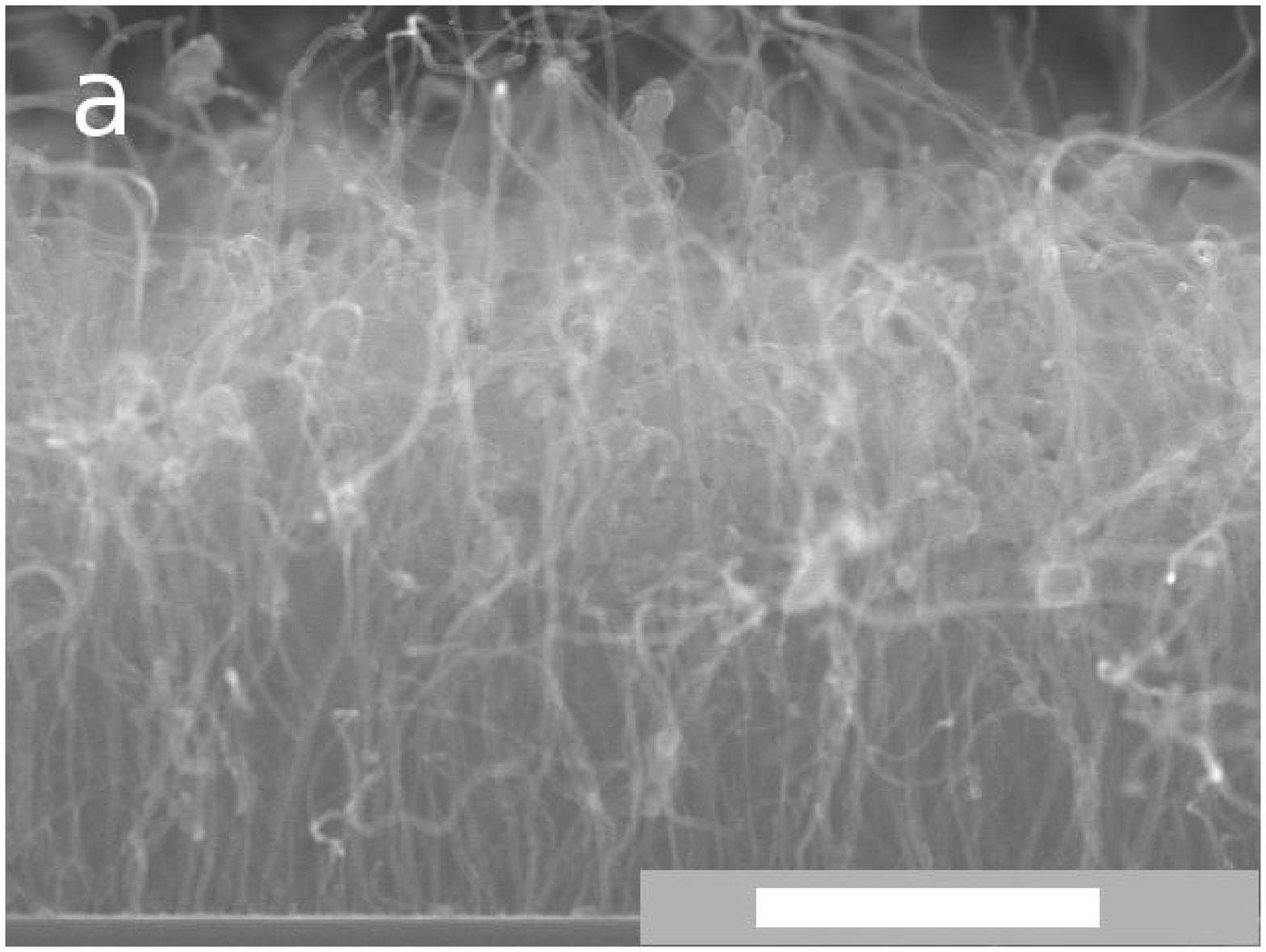}
\includegraphics[width=0.7\columnwidth]{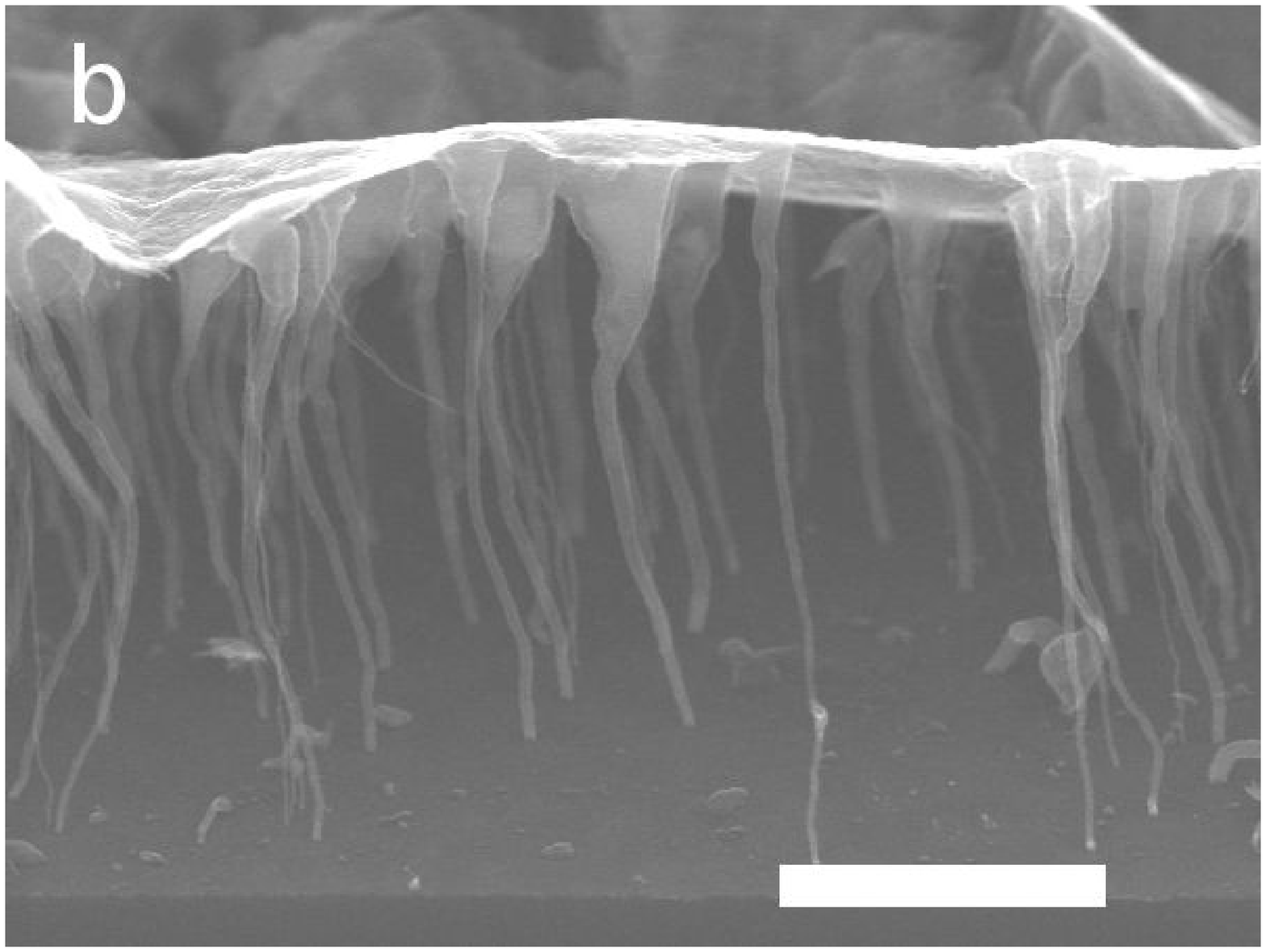}
\caption{\label{Figure1} Result of the CDV growth at 550 $^{\circ}$C during 30 minutes. a) Prior to growth, Argon was flown in the chamber during 30 minutes. 
Then propylene was introduced at 30 Torr. b) Propylene was introduced in the chamber without argon pretreatment. The scale bar is 1 $\mu m$ in both images.}%
\end{figure}

Figure 1 shows Scanning Electron Microscope (SEM) images of the product of Chemical Vapour Deposition (CVD) on a 6 nm thick nickel film. The growth was performed 
at 550 $^{\circ}$C in an Applied Material 200 mm DXZ chamber using propylene ($C_3H_6$) as carbon source. The Ni film was deposited by Physical Vapour 
Deposition on a 10 nm TiN layer deposited by CVD on a standard silicon 200 mm wafer. Depending on the pretreatment (initial step before the introduction 
of propylene) very different result are obtained. On one hand, with a pre-treatment consisting of an argon flow during 15 minutes, carbon nanotubes are 
obtained (Figure~\ref{Figure1}a). The nanotubes are multiwall with catalyst at the tip and about 10 nm in diameter (not shown). On the other hand, in 
absence of pre-treatment a complex nanostructure composed of a forest of nanotubes with a thin layer bonded on top of it is formed (Figure~\ref{Figure1}b).

A similar nanostructure was observed by Kondo and colleagues by CVD at 510  $^{\circ}$C using cobalt as catalyst and acetylene ($C_2H_2$) as carbon precursor.\cite{Kondo2008}
They proposed the following mechanism for the growth of this composite. When a substrate with a catalyst film is heated in 
presence of carbon precursor, the catalytic film first decomposes the source gas while remaining in its thin-film structure. This leads to the formation 
of few graphene layers on the catalyst. Second, the catalyst film breaks into particles. Third, multi-wall CNTs are synthesized from catalyst particles
 by the tip growth mode.

Multi- and single graphene layers are quite well known to form on Nickel films.\cite{Gamo1997,Yu2008,Reina2009,Pollard2009,Kim2009} Carbon nanotubes are also 
well known to grow from Nickel nanoparticles.\cite{Ren1999,Helveg2004,Hofmann2007} It is therefore surprising that the composite has not yet been reported on Ni. This fact, 
together with our observations already indirectly confirms the scenario proposed by Kondo and colleagues.\cite{Kondo2008} Indeed, in the CNT community, very thin Ni layers 
(1 nm or less) are dewetted into small droplets that subsequently catalyze the nanotube growth. The thinner the film the smaller the particles and therefore the
 thinner the nanotubes. This is important because the goal here is to synthesize the thinnest nanotubes which are more likelly to show interesting effects such 
as quantum confinement.\cite{Frank1998} 
\begin{figure}
\includegraphics[width=0.7\columnwidth]{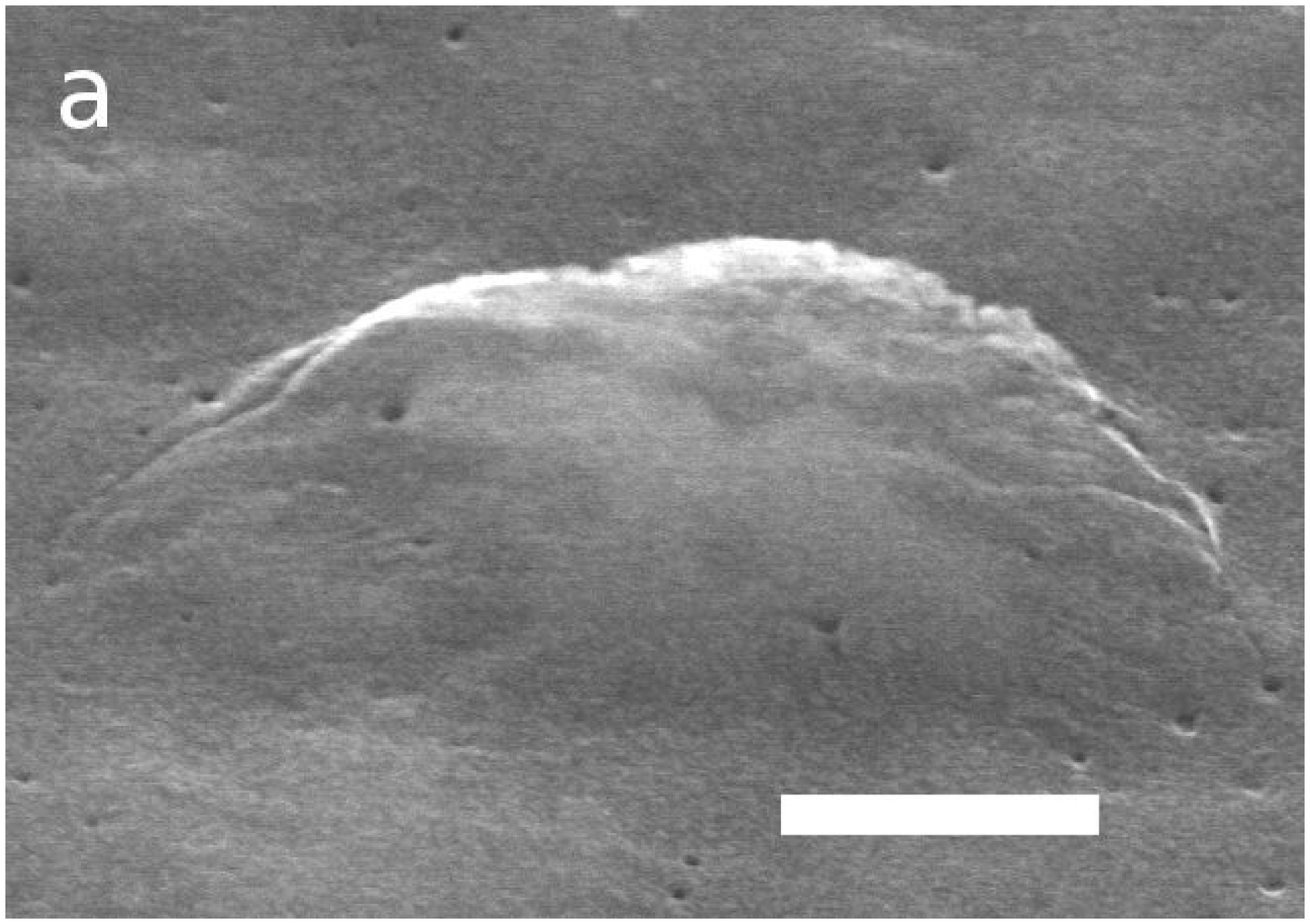}
\includegraphics[width=0.7\columnwidth]{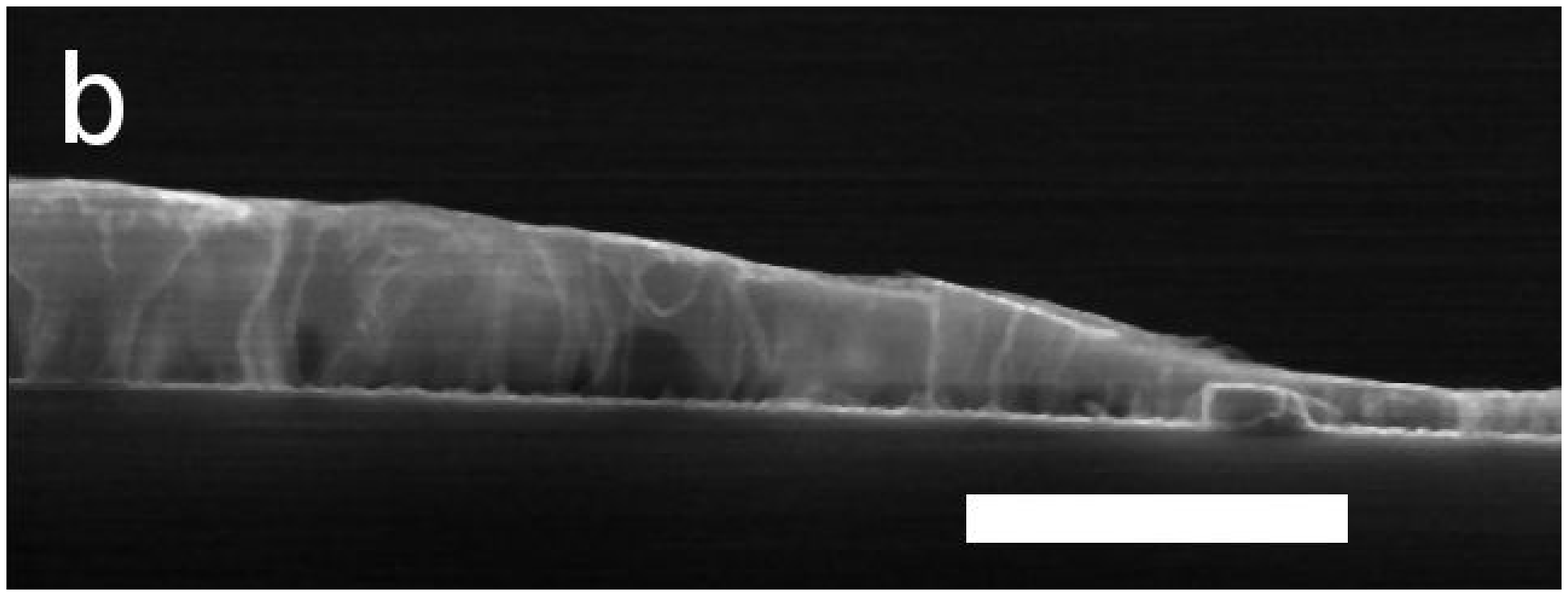}
\caption{\label{Figure2} Initial stage of the composite growth. The images are taken after growth in the same condition as in Fig.~\ref{Figure1}a but the 
growth lasted 5 minutes instead of 30 minutes.  a) Top view of the sample showing a bump below which CNT growth has started. b) Side view of such bump illustrating 
the presence of nanotubes.}%
\end{figure}
On the contrary, in the graphene community relativelly large Ni thicknesses (typically larger than 100 nm) are used since one wants to avoid a dicontinuous 
graphene layer due to dewetting. We used a catalyst layer of intermediate thickness (few nanometre) which may represent a better 
compromise between the necessity to keep the catalyst film intact during multilayer-graphene growth and the necessary dewetting after some time to start 
nanotube growth. 

This is comfirmed by the observation that when the Nickel film is intentionnaly dewetted using argon pre-treatment only nanotubes are formed. This
is also confirmed by the SEM images taken at the initial stages of the growth. After 5 minutes of growth under the same conditions as in Figure \ref{Figure1}a
there is an extended carbon film at the surface of the sample. In some places the catalyst has already dewetted and the CNTs growth has started. This is evidenced
by bumps at the surface and cracks from wich it is possible to see nanotubes below the film (See Figure \ref{Figure2}). Therefore our observation clearly establishes that the 
few graphene layers form first.

\begin{figure}
\includegraphics[width=0.49\columnwidth]{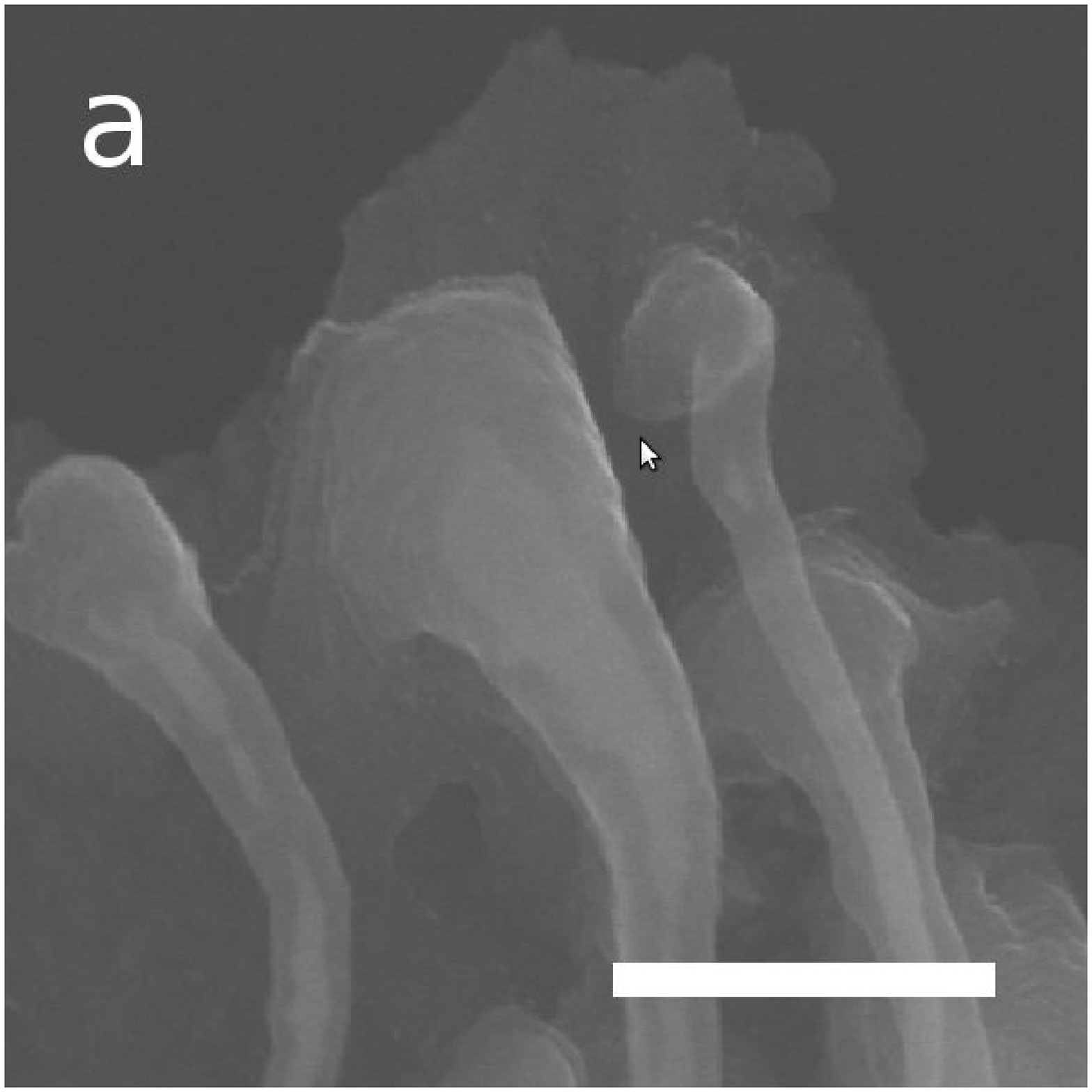}
\includegraphics[width=0.49\columnwidth]{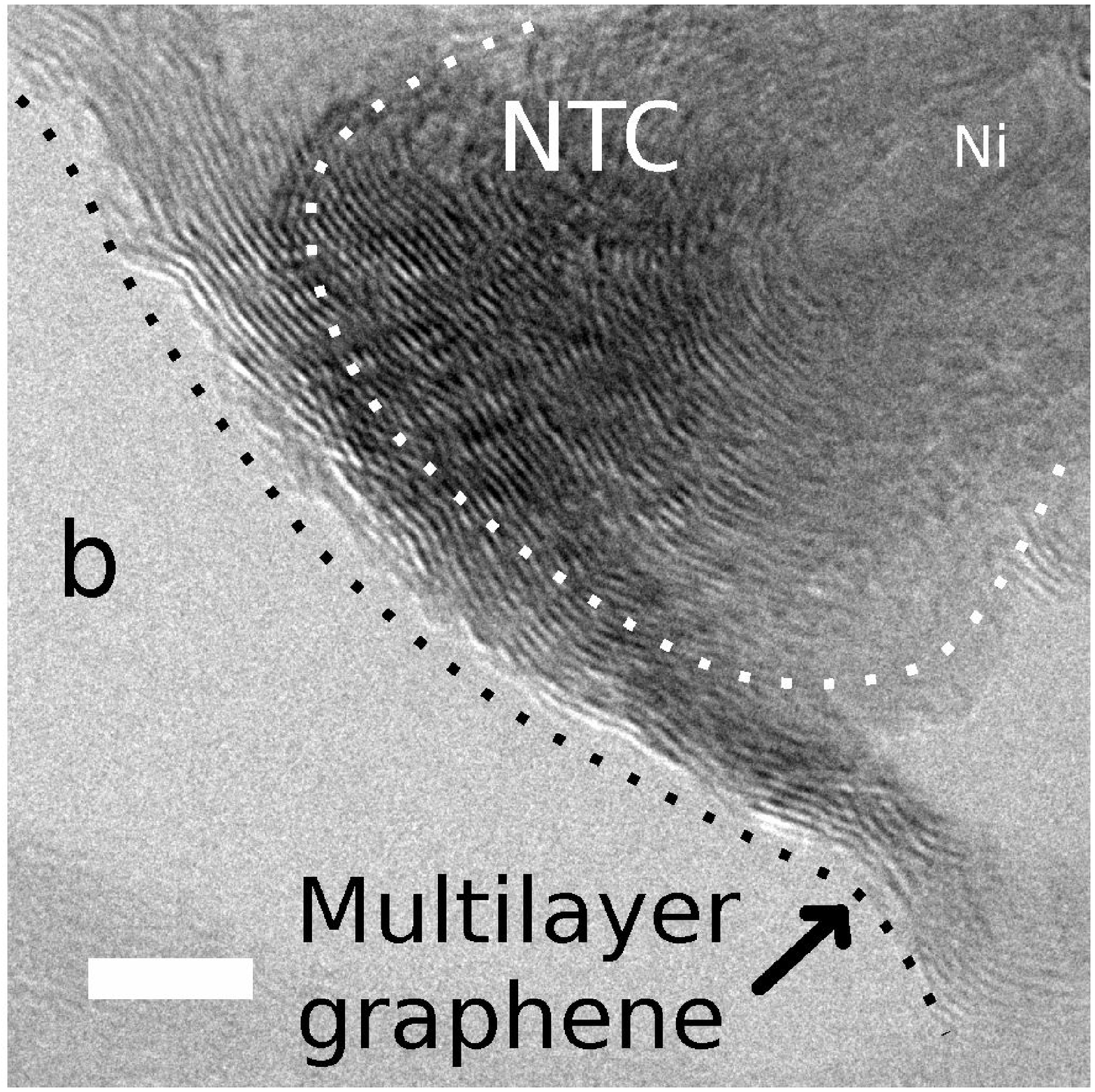}
\includegraphics[width=0.49\columnwidth]{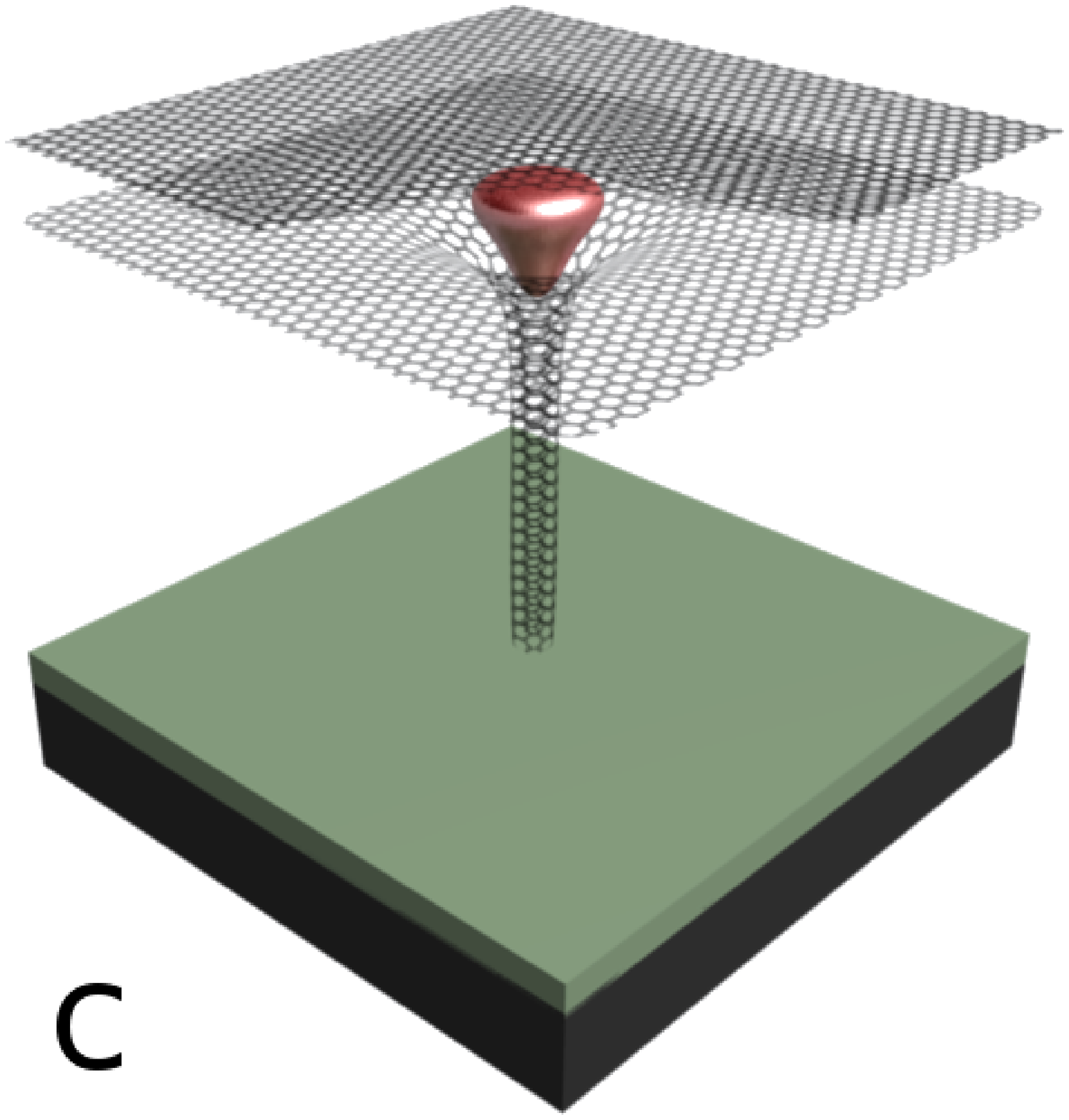}
\includegraphics[width=0.49\columnwidth]{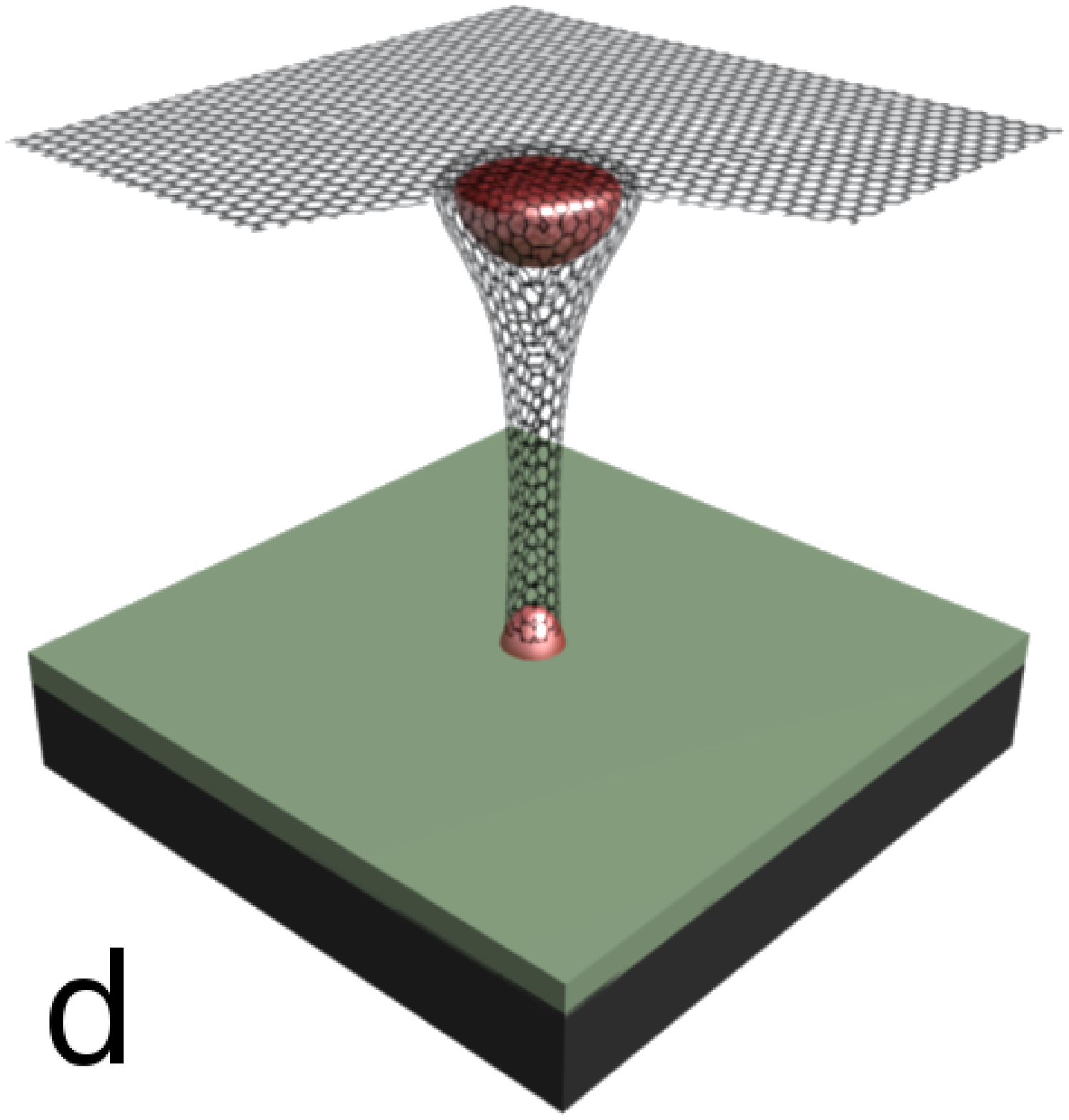}
\caption{\label{Figure3} a) Connexion between nanotubes and multilayer-graphene viewed from below by SEM. Ni catalyst can be seen in the NTCs. The scale bar is 300nm. 
b) HRTEM image of the carbon-nanotube/multilayer-graphene interface. The dotted lines are guides for the eyes to localize the nanotube tip and the multilayer graphene sheet. 
The scale bar is 5 nm. c) and d) Sketch of the two possible arrangement for the CNT/multilayer-graphene interface. The red particles represent the catalyst. 
Our result indicate that the interface corresponds to the configuration depicted in d) and that the nanotubes grow from their root.}%
\end{figure}

Figure \ref{Figure3} presents SEM and High Resolution Transmission Microscope (HRTEM) image of the junction between a nanotube and the graphene layers.The composite has 
been collected on an amorphous carbon TEM grid for HRTEM investigations. The nanotube are epitaxial under less than 10 layers of graphene. 
In the previous study,\cite{Kondo2008} there were indications that the nanotube originated from holes in the first graphene layer forming a 
trumpet like structure (Fig.~\ref{Figure3}c). Figure~\ref{Figure3}a clearly illustrates that in our case closed nanotubes are in fact expitaxially formed under the 
graphene layers (Sketched in Fig.~\ref{Figure3}d). The shape of this interface may therefore depend on the experimental conditions. This will have to be deeply investigated 
since it can have important consequenses concerning electrical contact between the nanotube and the graphene layers. Indeed, the contact resistance can be expected to 
be larger than in a configuration where the nanotube gradually transforms into a flat graphene sheet. Still, the contact electrical resistance should be reasonnable 
given that the graphene orientation in the nanotube and the multilayer graphene should match due to their epitaxial relation.\cite{Paulson2000} This will have to be studied by 
electrical measurements. The fact that the nanotube is closed by a relativelly important number of graphene layers suggests that NTCs grow from the root and not from their tip. 
Indeed, growth stops whenever the catalyst is completelly encaplusated.\cite{Helveg2004} The reduction of their diameter towards the substrate can be attributed to 
the loss of catalysing material durning the catalyst elongation/contraction process occuring during CNT growth.\cite{Helveg2004}

\begin{figure}
\includegraphics[width=0.7\columnwidth]{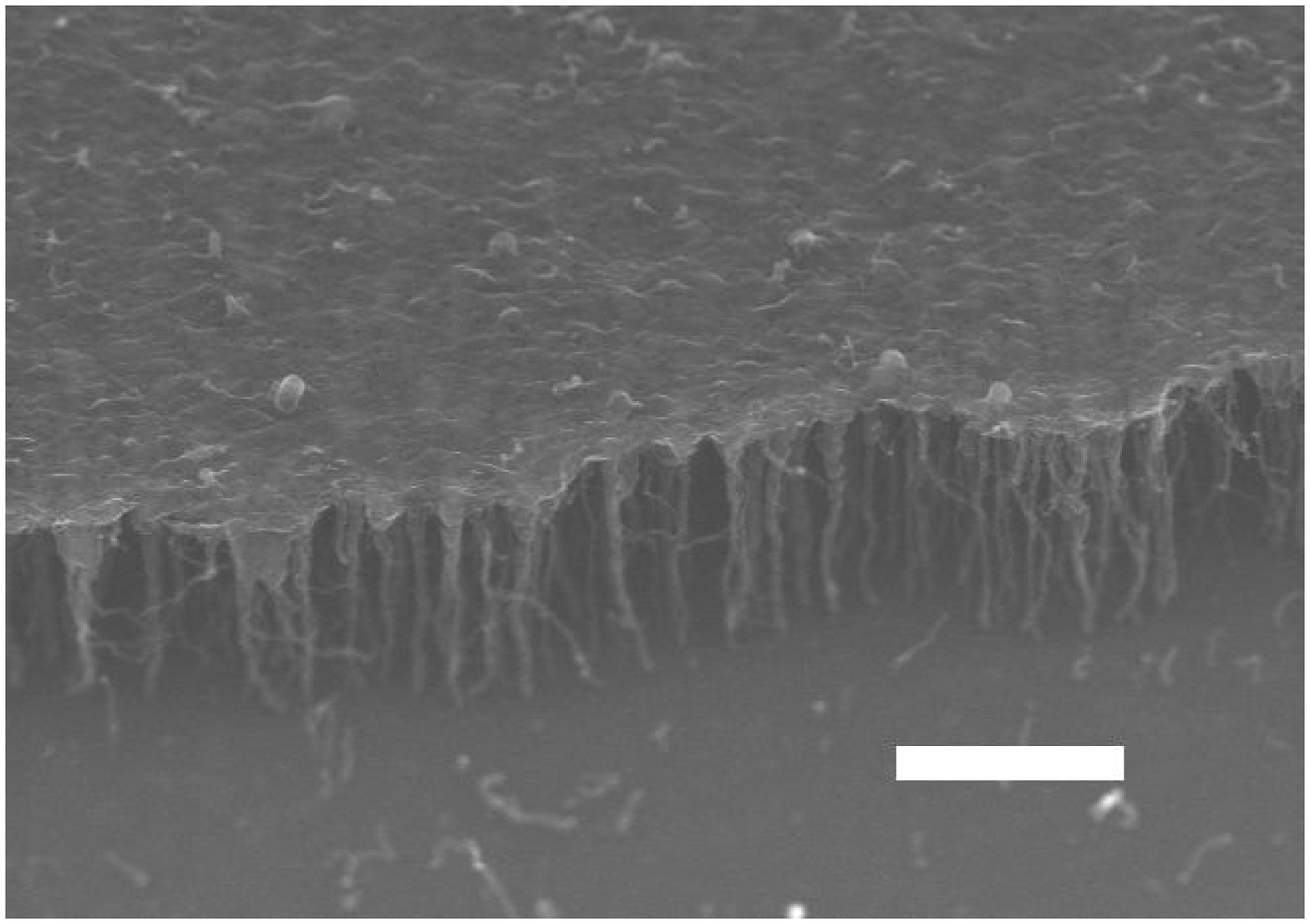}
\includegraphics[width=0.7\columnwidth]{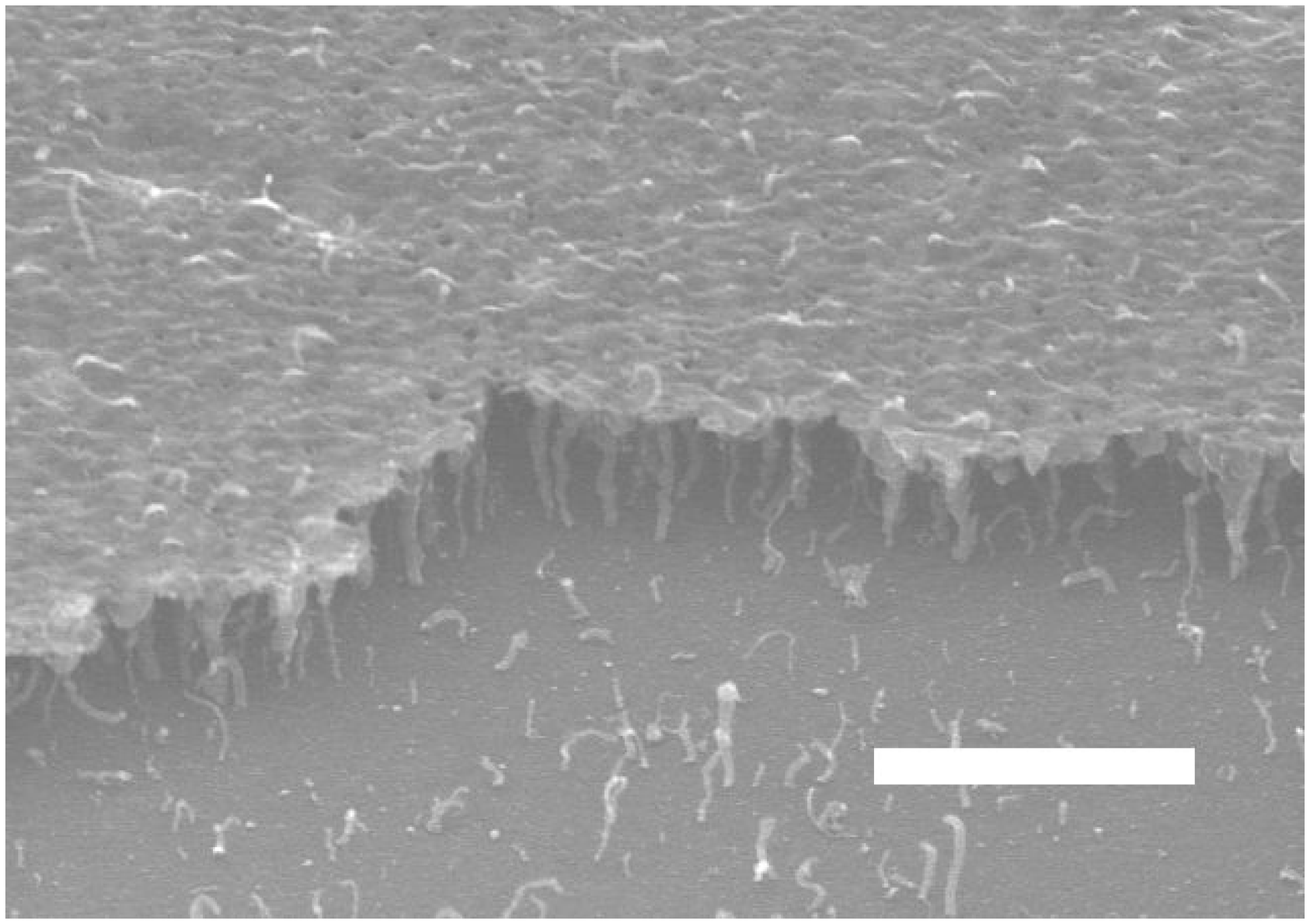}
\caption{\label{Figure4} Impact of the deposition temperature on the composite growth. a) Growth at 450 $^{\circ}$C. b) growth at 400 $^{\circ}$C.
In both cases a 15 minutes pretreatment with Argon was made prior the growth under 15 Torr of $C_3H_6$ during 15 minutes. The scale bar is 600 nm for both images}%
\end{figure}

Some key points in this mechanism remain unclear. In particular, why does the catalyst film brakes into particles
after some time? In the context of CNTs, it is established that Nickel atoms exhibit fast self-diffusivity in presence of graphene interface.\cite{Helveg2004,Hofmann2007} 
At the initial stage of CNT formation, graphene layers grow at the catalyst step edges and catalyst atoms rapidly move toward the graphene free region. This causes an 
elongation of the catalyst until the energy gained when binding the graphitic fibre to the Ni surface no longer compensate the increase in the Ni surface energy.
At this moment, the catalyst collapses into a more or less round particle.\cite{Helveg2004} We believe an analogous rapid diffusion of nickel atoms occurs in our case 
leading to the formation of catalyst particles, although we do not have direct evidences. Another key question concerns the way the organic precursor is provided to 
the catalyst after the graphitic film is formed. Indeed, graphene is a good barrier to gas diffusion\cite{Bunch2008} and nanotube CVD growth stops when the catalyst 
particle is encaplusated.\cite{Helveg2004} These questions will have to be investigated by \textit{in-situ} measurements to fully determine the detailed growth process.

We have also investigated the possibility to grow such type of composite at lower temperatures. This point is critical in the case of the co-integration of 
these carbon structures with CMOS architecture at the interconnection levels since the maximum temperature acceptable during the device fabrication is 450 $^{\circ}$C. 
As can be seen on Figure~\ref{Figure4}, composites are also obtained at temperatures as low as 400 $^{\circ}$C. Note that at these temperatures, dewetting does not occur as 
easily as at 550 $^{\circ}$C and the initial pretreatment with argon is necessary to achieve the composite. This is not surprising since Ni self-diffusion should be 
slower at lower temperatures. As expected, the lower the temperature, the shorter the CNTs. It appears that the quality of the multi-layer graphene film 
degrades with decresing temperature as seen from the number of visible defects on the SEM pictures.


VTR acknowledges the support from Nanoscience fundation of Grenoble. 
 


%
%

%


\end{document}